# STATUS OF THE IOTA PROTON INJECTOR*

D. Edstrom Jr.[†], A. Romanov[‡], D. Broemmelsiek, K. Carlson, J.-P. Carneiro, H. Piekarz,
A. Shemyakin, A. Valishev, Fermi National Accelerator Laboratory, Batavia, IL, United States

## Abstract

The IOTA Proton Injector (IPI), currently under installation at the Fermilab Accelerator Science and Technology facility (FAST), is a machine capable of delivering 20 mA pulses of protons at 2.5 MeV to the Integrable Optics Test Accelerator (IOTA) ring. First beam in the IPI beamline is anticipated in the first half of 2024, when it will operate alongside the existing electron injector beamline to facilitate further beam physics research and continued development of novel accelerator technologies at the IOTA ring. This report details the expected operational profile, known challenges, and the current state of installation.

## INTRODUCTION

The proton injector for the Integrable Optics Test Accelerator (IOTA, [1]) is an important capability for the research program at FAST/IOTA [2]. IOTA is a research storage ring constructed and operated by Fermilab to study advanced concepts in accelerator physics. At present, the studies at IOTA are performed with electron beams of up to 150 MeV energy that has negligible space charge effects. The commissioning of 2.5 MeV IOTA proton injector with a beam current of 20 mA will allow experiments with intense beams with strong space charge tune shifts that constitute $10 - 15$ % of the total betatron phase advance in the ring. The rich menu of intense-beam research experiments planned at IOTA can be served with either electron or proton beams. The main research thrusts and the corresponding type of beams are:

- Demonstration of integrable optics with special nonlinear magnets (e, p)
- Demonstration of integrable optics with octupoles (e, p)
- Demonstration of integrable optics with non linear electron lenses (e, p)
    - Thin radial kick of McMillan type
    - Axially symmetric kick in constant beta function
- Space-charge compensation with electron lenses (p)
- Space-charge compensation with electron columns (p)
- Optical stochastic cooling (e)
- Electron cooling (p)
    - Electron cooling of protons
    - Diagnostics through recombination
    - Electron cooling and nonlinear integrable optics

## PROTON SOURCE

Protons are to be generated with the kinetic energy of 50 keV using a Duoplasmatron ion source (IS) configured for proton production as described in Fig. 1. This source was previously used as a part of the High Intensity Neutrino Source (HINS) experiment [3], and was subsequently adopted by the FAST facility as it fits the desired beam parameters for proton production in the IPI beamline, as shown in Fig. 2. The IS was resurrected briefly in site in the HINS enclosure at the Meson Detector Building (MDB) before being fully disassembled and cleaned. In the process of disassembly, it was noted that the extraction aperture of 1.25 mm in use with HINS was unusually large for a duoplasmatron source, and has been replaced with a 0.7 mm aperture. It has since been reassembled at the FAST facility.

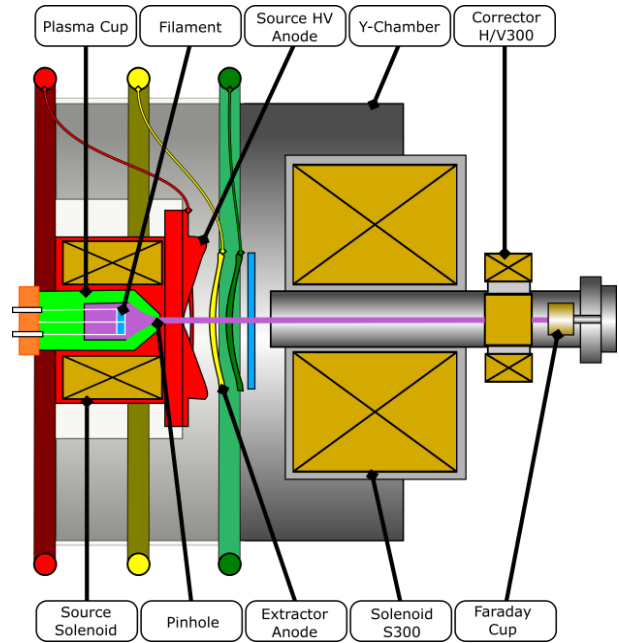

Figure 1: A schematic view of the IS in a test configuration. For extraction of 50 keV protons, the source high voltage anode is charged to a static $+50$ kV, the extractor anode is set to 40 kV, and the green anode is at ground potential. The S300 solenoid is the first element of the LEBT.

Filaments are prepared on-site using a mixture of Barium, Calcium, and Strontium Carbonates along with Isoamyl Acetate to assist in binding the mixture to the filament. The mixture is dissolved in Acetone to form a thin solution and the cleaned Nickel filament is dipped to deposit a modest coating of the carbonate mixture. It is then activated in a filament activation station referred to simply as *the Jar*, as shown in Fig. 3, by slowly increasing the filament current to the nominal 21 A, as seen in Fig. 4. This burns away the nitrocellulose binder and ultimately converts the carbonate mixture into oxides [4], lowering the work function and facilitating electron transmission on the filament surface. This is checked at points by applying a bias voltage between the filament and a plate connected to ground. We then flash the filament by raising the filament supply current to 22 A for short periods and check the work function by monitoring the current draw from the HV supply.

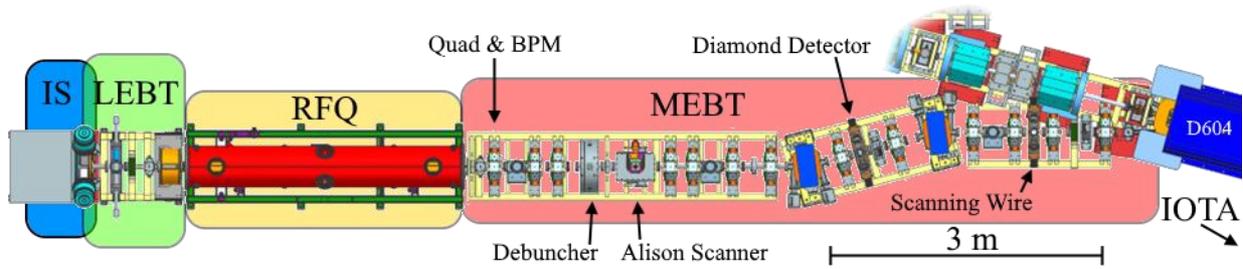

Figure 2: The IPI beamline meets the existing electron injector beamline at the bend dipole D604 and then injects into the IOTA ring along the same trajectory as electrons at 100 to 150 MeV. As such, time must be taken between electron and proton runs in the IOTA ring to allow for the necessary configuration changes.

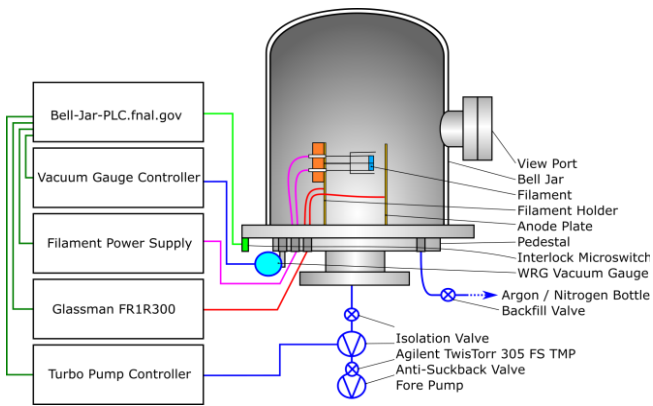

Figure 3: The filament activation station or simply the Jar, named for the large, evacuated bell jar in which activation is performed.

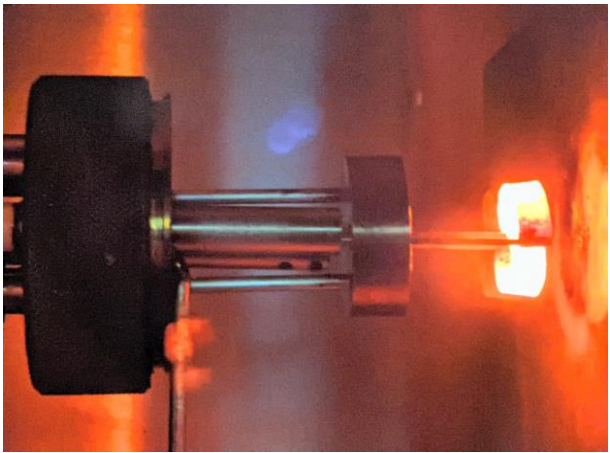

Figure 4: A filament assembly mounted in the Jar during the activation.

Table 1: Selected Parameters Specified for or Expected from the IPI and the IOTA Ring Under Proton Operation

| | Parameter | Nom. | Unit |
|---|---|---|---|
| LEBT | Energy | 50 | keV |
| | Proton beam current | 20 | mA |
| | Pulse length (99 %) | 350 | μs |
| | Source pulse rate | 1 | Hz |
| | Transverse beam size | 0.7 | mm |
| MEBT | Energy | 2.5 | MeV |
| | RF pulse rate | 1 | Hz |
| | RFQ frequency | 325.0±0.5 | MHz |
| | RFQ duty factor | < 0.002 | % |
| | Phase/Amp. stability | 1°/1 % | |
| | Beam pulse length | 2 | μs |
| | Bunch length (1$\sigma$) | 0.3 | ns |
| IOTA with protons | Proton beam energy | 2.5 | MeV |
| | Relativistic $\beta$ | 2.66 · 10$^{-3}$ | keV |
| | Circumference | 40 | m |
| | Proton RF frequency | 2.19 | MHz |
| | Revolution period | 1.83 | μs |
| | RF voltage | 1 | kV |
| | Geometric emittance | 3.5 | μm |
| | $\Delta p/p$ (RMS) | 0.07 | % |
| | Beam current | 8 | mA |
| | Momentum compaction | 0.07 | |
| | Betatron tunes ($Q_x$, $Q_y$) | 5.3, 5.3 | |

## LEBT

The IPI LEBT, as seen in Fig. 5, is a short section of beamline designed to transport beam from the IS to the RFQ. Features of note are the two solenoids (SOL) that provide focusing to allow for injection into the RFQ, a gate valve (GV) to isolate the IS from the RFQ, a toroid (TOR) to monitor beam current from the source, and an electrically isolated diaphragm (EID). The EID has two primary roles. The first is to measure the intensity of the beam intercepted, allowing for tuning of the solenoids to maximize beam transmission to the RFQ or to perform other beam studies. The second goal is to assist with beam space charge neutralization. A pair of beamline corrector packages (H/V) also provide basic horizontal and vertical beam steering capabilities. Finally, scrapers will also be installed for commissioning. These may be removed for operation as the vacuum chamber they occupy restricts the beam aperture. The simulations shown in Fig. 5 are performed with no space charge effects.

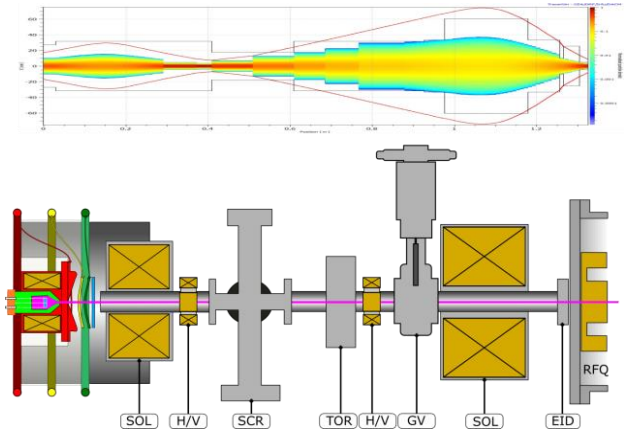

Figure 5: The LEBT configuration and a TraceWin simulation showing aperture restrictions and beam sizes and density through the LEBT.

## RFQ

Like the IS, the RFQ was also adopted from the HINS experiment and is tasked with accelerating the proton beam from 50 keV to 2.5 MeV [5, 6]. Due to a cooling-to-vacuum leak [7], it cannot currently be temperature-regulated. The low duty factor cited in Table 1 will assist in mitigating the effects of frequency shift, and a test performed in 2020 indicates that the net effect will be a $\sim$ 5.6 kHz drop in the resonant frequency over the first 8 hours for an average CW power of $\sim$ 100 W. It will level out as the cavity approaches equilibrium, and the initial plan will be to manage most of the frequency shift with the cavity tuner and the LLRF frequency. This was not a workable plan for the same RFQ when it was used as a part of the HINS experiment as the required duty factor was much higher and the LLRF frequency was fixed. To address quick frequency fluctuations, the FRQ will be operated in self-excited loop with the bunching cavity following it.

## MEBT

The IPI MEBT transports the proton beam from the RFQ through the D604 bend dipole, and into IOTA. The section starting from D604 is the existing 300 MeV Electron Injector beamline. Beam parameters at he exit of the RFQ were gathered from Ref. [6, Table 7 and Fig. 16] and later confirmed with simulations [8]. The MEBT includes the aforementioned bunching RF cavity [9] to reduce the beam momentum spread from about 0.5 % to 0.06 % in a linear approximation. Another important structure in the line is a dogleg that is necessary to excite and match dispersion at the injection point in IOTA. To obtain a fully matched beam while avoiding the aperture limitations, a total of 17 quadrupoles will be used, including 12 in the newly installed line and 3 additional Hallbach array quadrupoles based on permanent magnets fitted in the tight spots of the existing line to assist the strict matching criteria. Figure 6 shows RMS beam envelope for $3\sigma$ in both planes along with the corresponding physical aperture.

The MEBT line will be equipped with several pieces of instrumentation for monitoring the beam, including 5 BPMs, each nested into the profile of a beamline quadrupole, an MEBT Alison Scanner for measuring the emittance of the beam following the RFQ, and two scanning wire assemblies to measure transverse beam profiles. A diamond detector is slated for installation in the dogleg section of the beamline.

Beamline magnets include trim dipole packages, dipole bend magnets, and quadrupole magnets repurposed from the low energy Electron Injector beamline build-out. There are also two multi-function magnets, in which each of four coils are independently controlled to affect trim dipole and skew-quadrupole fields in a single element. These multifunction magnets are also used in the IOTA Ring.

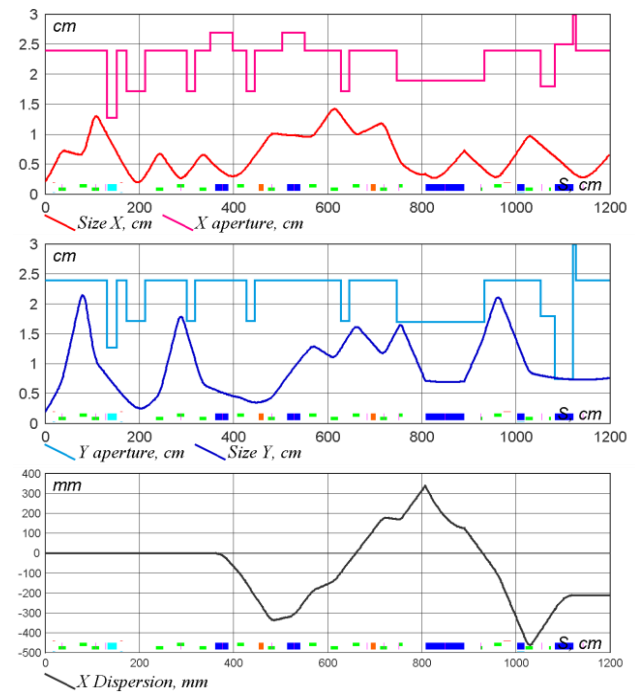

Figure 6: Beam envelope for $3\sigma$ in the horizontal (top) and vertical (middle) planes along with the corresponding aperture envelopes, and horizontal dispersion (bottom). Lattice starts from the exit of RFQ and extends to IOTA which starts around azimuth of 1150 cm.

## CURRENT STATUS

All instrumentation and beamline elements are in-hand. The acquisition of stands, cables, and other infrastructure is well underway with the goal of beginning the installation right after the end of the ongoing experimental run with elec-trons. The first proton beam in IOTA is expected at the beginning of 2024. The reassembled source has been charged up to 50 kV, and commissioning and characterization are pending coordination with radiation safety personnel.

Significant work remains to install and commission the IPI beamline, but success will provide the opportunity of investigating numerous novel accelerator technologies and concepts in the IOTA ring using 2.5 MeV protons.

## COMMISSIONING PLAN

We plan a staged approach to the commissioning of IPI. The first step would be to extract at least 20 mA of current from the ion source with the necessary pulse length of 350 µs. The long pulse length is necessary to get to a steady state with near complete space charge compensation in the LEBT line. At this stage LEBT will be terminated into the diagnostics stand equipped with various instruments such as: the Allison scanner; the Faraday cup; an electric deflector with scraper pads.

At the second stage, the LEBT line will be attached to the RFQ with a short section of the MEBT line ending with a Faraday cup after the Allison scanner. In this configuration we will maximize the RFQ transmission efficiency and test the interface between RFQ and bunching cavity.

The third stage would consist of the commissioning of the MEBT line up to the injection point into IOTA. The final stage will be dedicated to capturing and storing beam in the IOTA ring.

## ACKNOWLEDGEMENTS

We would like to acknowledge and thank all those who have worked so hard to make the IPI, the IOTA ring, and the Electron Injector beamline a reality, both within the FAST facility departments as well in the various organizations across Fermilab.

## REFERENCES

[1] S. Antipov *et al.*, "IOTA (Integrable Optics Test Accelerator): Facility and experimental beam physics program," *J. Instrum.*, vol. 12, no. 03, p. T03002, 2017. doi:10.1088/1748-0221/12/03/t03002

[2] M. Church *et al.*, "Proposal for an accelerator R&D user facility at Fermilab's Advanced Superconducting Test Accelerator (ASTA)," Fermilab, Batavia, IL, USA, Rep. FERMILAB-TM-2568, 2013. doi:10.2172/1422196

[3] W. M. Tan, "Characterization of the proton ion source beam for the high intensity neutrino source at Fermilab," PhD thesis, Indiana University, Bloomington, IN, USA, 2010.

[4] F. Rosebury, *Handbook of Electron Tube and Vacuum Techniques*. Addison-Wesley Publishing Company, Inc., 1965.

[5] G. V. Romanov and A. Lunin, "Complete RF Design of the HINS RFQ with CST MWS and HFSS," in *Proc. ICAP'09*, San Francisco, CA, USA, Aug.-Sep. 2009, pp. 340–342. https://jacow.org/ICAP2009/papers/THPSC047.pdf

[6] P. N. Ostroumov, V. N. Aseev, and A. A. Kolomiets, "Application of a new procedure for design of 325 MHz RFQ," *J. Instrum.*, vol. 1, no. 4, p. P04002, 2006. doi:10.1088/1748-0221/1/04/P04002

[7] T. N. Khabiboulline *et al.*, "Experiences with the Fermilab HINS 325 MHz RFQ," in *Proc. LINAC'10*, Tsukuba, Japan, Sep. 2010, pp. 515–517. https://jacow.org/LINAC2010/papers/TUP048.pdf

[8] TraceWin simulation package. http://irfu.cea.fr/Sacm/logiciels/

[9] G. V. Romanov *et al.*, "CW Room-Temperature Bunching Cavity for the Project X MEBT," in *Proc. PAC'11*, New York, NY, USA, 2011, pp. 1900–1902. https://jacow.org/PAC2011/papers/WEP221.pdf